\documentclass[fleqn,twoside]{article}
\usepackage{espcrc2}

\usepackage[figuresright]{rotating}

\newcommand{\AmS}{{\protect\the\textfont2
  A\kern-.1667em\lower.5ex\hbox{M}\kern-.125emS}}
\newcommand{\be}{\begin{equation}}
\newcommand{\ee}{\end{equation}}

\title{Bhabha Scattering at NNLO \thanks{Work supported by the SNF under contract 200020-117602.
Z\"urich preprint number ZH-TH 06/08.}}

\author{R. Bonciani\address[MCSD]{Institut f\"ur Theoretische Physik,
        Universit\"at Z\"urich,
      CH-8057, Z\"urich, Switzerland }%
     and
       A. Ferroglia \addressmark
}
\begin{document}

\begin{abstract}
We review the status of the calculation of next-to-next-to-leading order corrections to
large angle Bhabha scattering in pure QED. After discussing the electron-loop and
photonic corrections, we focus on the recently calculated two-loop virtual corrections
involving a heavy-flavor fermion loop. We conclude by assessing the numerical impact of
these corrections on the Bhabha scattering cross section at colliders
operating at a center of mass energy of about $1$-GeV.
\vspace{1pc}
\end{abstract}

\maketitle

\section{INTRODUCTION}

The scattering process $e^- \, e^+ \to e^- \, e^+$ is named after
the author that first calculated its cross section in 1936
\cite{Bha}. Nowadays, Bhabha scattering is of crucial importance
in the study of elementary particle phenomenology, since it is the
process chosen for the luminosity determination at all $e^+  e^-$
colliders. In particular, at colliders operating at a center of
mass energy $\sqrt{s} \sim 1-10$ GeV, the luminosity is measured
by observing Bhabha events at large scattering angles. In this
kinematic  region, the Bhabha scattering cross section is large
and QED dominated: these two properties  allow precise
experimental measurements and an accurate  theoretical evaluation
of the cross section in perturbation theory.
Any realistic calculation of the theoretical Bhabha scattering cross section must take
into account the detector geometry and experimental cuts. For this purpose a number of
sophisticated MC generators were developed
\cite{BHAGENF,BABAYAGA,BABAYAGA-1,BHWIDE,MCGPJ}. In particular, at KLOE/DA$\Phi$NE, the
experimental accuracy reached on the measurement of Bhabha scattering events, joined
with the theoretical accuracy  of the MC generator {\tt BABAYAGA}
\cite{BABAYAGA,BABAYAGA-1,CCCphipsi08}, allowed to determine the luminosity with an
error of $0.1 \%$ \cite{Alo}.
Since the theoretical error on the Bhabha scattering differential
cross section directly affects the luminosity measurement
precision, in recent years a significant effort was devoted to the
calculation of  perturbative corrections to this process. The
next-to-leading order corrections are well  known even within the
full Standard Model \cite{Bhabha1loop}. At the
next-to-next-to-leading order (NNLO), for what concerns the
electroweak corrections, only the logarithmically enhanced terms
were calculated \cite{KA}. In pure QED, instead, the situation is
sensibly different.
The first diagrammatic calculation of the  two-loop QED virtual corrections to Bhabha
scattering can be found in \cite{Bern}. However,  this result was obtained
by employing dimensional regularization (DR) to regulate both soft and collinear
divergencies, while all the  available MC generators employ the electron mass $m_e$ as a
collinear regulator.  Today, the complete set of NNLO corrections to Bhabha scattering
in pure QED have  been evaluated using $m_e$ as a collinear regulator. The two-loop
Feynman diagrams involved in the calculation can be divided in three gauge independent
sets: {\em i)} diagrams without fermion loops (``photonic'' diagrams), {\em ii)}
diagrams involving a closed electron loop, and {\em iii)} diagrams involving a closed
loop of a fermion heavier than the electron. In the following three sections  we
consider the calculation of each set of corrections separately. In
Section~\ref{Conclusions}, we conclude by briefly discussing the numerical impact of
these corrections on the Bhabha scattering cross section at $\sqrt{s}
\sim 1$ GeV.

\section{PHOTONIC CORRECTIONS}

All the NNLO photonic corrections, with the exception of the ones arising  from the
two-loop photonic boxes, can be obtained in a straightforward way, retaining the full
dependence on $m_e$ \cite{BF}, by using a technique that is by now  standard in
multi-loop calculations. This technique is based on the Laporta algorithm
\cite{Laportaalgorithm}  for the reduction of the Feynman diagrams to the Master
Integrals (MIs), and then on the differential equation method \cite{DiffEq}  for their
analytic evaluation.
However, a calculation of the two-loop photonic boxes retaining the full dependence on
$m_e$ is beyond the reach of this technique. This is due to the fact that the number
of Master Integrals belonging to the same topology after the reduction process is
large\footnote{Some topologies have up to 6 MIs.}.
This means that
one should be able to solve analytically large systems of  first-order
ordinary differential equations; this is in general not possible.
Alternatively, one could use the Mellin-Barnes techniques \cite{Mellin}
to calculate the MIs one by one\footnote{Partial results can be found in 
\cite{Smirnov,Kalm}}.
 The
calculation, however, is very complicated and, moreover, from a phenomenological point
of view, to keep the full dependence on the electron mass is not important.
In fact, the physical problem exhibits a well defined mass hierarchy. The mass of the
electron is always very small compared to the other kinematic invariants and it
can be safely neglected everywhere, with the exception of the terms in which it acts as a collinear
regulator.
It can be shown that the collinear structure of the NNLO corrections is the following
\be
\frac{d \sigma^{(NNLO)}}{d \sigma^{(\mbox{\tiny{Born}})}}  =
\frac{\alpha^2}{\pi^2} \sum_{i=0}^{2} \delta^{(i)} L_e^i +
\mathcal{O}\left(\frac{m_e^2}{s}, \frac{m_e^2}{t}\right) \, ,
\label{exp-photonic}
\ee
where $L_e=\ln{(s/m_e^2)}$ and where the coefficients $\delta^{(i)}$ are functions of
the scattering angle $\theta$ and, in general, of the mass of the heavy fermions
involved in the virtual corrections. For any practical purpose, the approximation given
by Eq.~(\ref{exp-photonic}) is sufficient for a phenomenological description of the
process\footnote{It can be shown that the terms
suppressed by a positive power of $m_e^2/s$ do not play any phenomenological role
already at very low c.m. energies, $\sqrt{s} \sim 10$ MeV. Moreover, the terms $m_e^2/t$
(or $m_e^2/u$) become important in the very forward (backward) region, unreachable for
the experimental set ups. }.
%
In the photonic case, the coefficients of the squared and single collinear logarithm in
Eq.~(\ref{exp-photonic}) were obtained in \cite{L2,Bas}. The precision required for
luminosity measurements  at $e^+ e^-$ colliders demanded the calculation of the
non-logarithmic coefficient, that was  obtained in \cite{Pen} through the {\em infrared
matching} to the massless  approximation. The technique of \cite{Pen} allowed to
reconstruct the photonic differential cross section in the $s \gg m_e^2 \neq 0$ limit
from the calculation in \cite{Bern}, where $m_e$ was set to zero from the  start. The
method employed in \cite{Pen} involves a change of   regularization scheme for the
collinear divergencies originating from a  vanishing electron mass. A method based on a
similar principle was subsequently  developed in \cite{Mitov:2006xs,BecMel}; the authors
of \cite{BecMel} confirmed  the result of \cite{Pen} for the NNLO photonic corrections
to the Bhabha  scattering differential cross section.


\section{ELECTRON LOOP CORRECTIONS}

The NNLO electron loop corrections arise from the interference of two-loop  Feynman
diagrams with the tree-level amplitude as well as from the interference of one-loop
diagrams, as long as one of the diagrams contributing to each term involves a closed
electron loop. This set of corrections   presents a single two-loop box topology, and it
is therefore technically less challenging to evaluate  with respect to the photonic correction
set.  The calculation of the electron loop corrections was completed a few years ago
\cite{electronloop}; the  final result retains the full dependence of the differential
cross section on  the electron mass $m_e$.  The Master Integrals involved in the
calculation were identified by means of the Laporta algorithm and evaluated with the
differential equation method.   As expected, after UV renormalization the differential
cross section presented only  residual IR  poles that were removed by adding the
contribution of the  soft photon emission diagrams.  The resulting NNLO differential
cross section could be conveniently written in terms of  1- and 2-dimensional Harmonic
Polylogarithms (HPLs) of maximum weight three \cite{HPLs}.
Expanding the cross section in the limit $s, |t| \gg m_e^2$, the ratio of the NNLO
corrections to the Born cross section can be  written as in Eq.~(\ref{exp-photonic}):
\be
\frac{d \sigma^{(2,\mbox{\tiny{EL}})}}{d \sigma^{(\mbox{\tiny{Born}})}}  =
\frac{\alpha^2}{\pi^2} \sum_{i=0}^{3} \delta^{(\mbox{\tiny{EL},i})} L_e^i +
\mathcal{O}\left(\frac{m_e^2}{s}, \frac{m_e^2}{t}\right) \, .
\label{exp-electron}
\ee
The explicit expression of all the coefficients $\delta^{(\mbox{\tiny{EL},i})}$,
obtained by expanding the results of \cite{electronloop} was confirmed by two different
groups  \cite{BecMel,Act}. It is easy to check that the cubed collinear logarithm in
Eq.~(\ref{exp-electron}) cancels against the corresponding  term arising from the soft
pair production graphs \cite{Arbuzov:1995vj}.

\section{HEAVY FLAVOR CORRECTIONS}

Finally, we consider the corrections originating from two-loop Feynman
diagrams involving a heavy flavor fermion loop\footnote{In this context,
and at $\sqrt{s} \sim 1$ GeV, by heavy flavor fermion we mean muon and tau
leptons, as well as $b$- and $c$-quarks. Top quarks completely decouple at
intermediate energies.}. Since this set of corrections
involves one more mass scale with respect to the corrections analyzed in the
previous sections, a direct diagrammatic calculation is in principle a more
challenging task.
Recently, Becher and  Melnikov applied their technique based on SCET to Bhabha
scattering and obtained the heavy flavor NNLO corrections in the limit in which
$s, |t|, |u| \gg m_f^2 \gg m_e^2$, where $m_f^2$ is the mass of the heavy
fermion running in the loop \cite{BecMel}. Their result was very soon confirmed
by means of a method based on the asymptotic expansion of Mellin Barnes
representation of the Master Integrals involved in the calculation \cite{Act}.
However, the results obtained in the approximation
$s, |t|, |u| \gg m_f^2 \gg m_e^2$ cannot be applied to the case in which the
$\sqrt{s} < m_f$ (as in the case of a tau loop at $\sqrt{s} \sim 1$ GeV), and
they apply only to a relatively narrow angular region perpendicular to the
beam direction when $\sqrt{s}$ is not very much larger than $m_f$ (as in the
case of top quark loops at ILC). It was therefore necessary to calculate the
heavy flavor corrections to Bhabha scattering assuming only that
$s, |t|, |u|, m_f^2 \gg m_e^2$.
The technical problem can be simplified by considering carefully, once
more, the structure of the collinear singularities of this set of corrections.
The ratio of the NNLO heavy flavor corrections to the Born cross section is
given by
\be \label{exp-heavyflavor}
\frac{d \sigma^{(2,\mbox{\tiny{HF}})}}{d \sigma^{(\mbox{\tiny{Born}})}}  =
\frac{\alpha^2}{\pi^2} \sum_{i=0}^{1} \delta^{(\mbox{\tiny{HF},i})} L_e^i +
\mathcal{O}\left(\frac{m_e^2}{s}, \frac{m_e^2}{t}\right) \, .
\ee
It is possible to prove that, in a physical gauge, all the collinear
singularities factorize and can be absorbed in the external field
renormalization \cite{FreTay}. This observation has two consequences
in the case at hand. The first one is that box diagrams are free of
collinear divergencies in a physical gauge; since the sum of all boxes forms a
gauge independent block, it can be concluded that
the sum of all box diagrams is free of collinear
divergencies in any gauge. The second consequence is that the single collinear
logarithm in Eq.~(\ref{exp-heavyflavor}) arises from vertex corrections only.
Moreover, if one chooses on-shell UV renormalization conditions,
the irreducible two loop vertex graphs
are free of collinear singularities.
%
%
Therefore, among all the two-loop diagrams contributing to the
NNLO heavy flavor corrections to Bhabha scattering, only the
reducible vertex corrections are logarithmically divergent in the
$m_e \to 0$ limit\footnote{Additional collinear logarithms arise
also from the interference of one-loop vertex and self-energy
diagrams.}, and they are very easy to calculate even if they
depend on two different masses.
By exploiting these two facts, we were recently able to obtain the NNLO heavy  flavor
corrections to the Bhabha scattering differential cross section \cite{hfbha,hfbhaJHEP},
assuming only that $s, |t|, |u|, m_f^2 \gg m_e^2$. In particular, in obtaining
the analytic expression for the NNLO cross section,
we worked in the Feynman gauge, setting $m_e =0$ from the
start in all the diagrams with the exception of the reducible ones
and of the interference of one-loop graphs.
This procedure allowed us to effectively eliminate a mass scale from the
two-loop boxes, so that the evaluation could be carried out with the techniques
already employed in the diagrammatic calculation of the electron loop
corrections\footnote{The necessary MIs can be found in\cite{hfbhaJHEP,MIsHeavy}}.
%
%
We want to stress that in this approach individual box diagrams
are singular in the $m_e \to 0$ limit and the collinear
singularities appear as additional poles in the dimensional
regulator $\epsilon$; however it is easy to prove that such
divergencies cancel in the sum of all the box diagrams.
%
%
%
%
%
%
By expanding the analytic results of \cite{hfbha,hfbhaJHEP} it was possible to
check the heavy flavor cross section in the $s,|t|,|u| \gg m_f^2 \gg m_e^2$
limit, which was previously known \cite{BecMel,Act}.
At intermediate energy colliders like DA$\Phi$NE,
the exact dependence on $m_f$ of the
results of \cite{hfbha,hfbhaJHEP} allows to account for the contribution of
muons, taus, $b$- and $c$-quark loops to the Bhabha scattering
cross section. In the case in which the heavy flavor fermion is a quark, it was
straightforward
to modify the calculation of the two-loop self-energy diagrams to obtain the
mixed QED-QCD corrections to Bhabha scattering \cite{hfbhaJHEP}.
An alternative numerical approach to the calculation of the heavy flavor
corrections to Bhabha scattering, based on dispersion relations, was pursued in
\cite{Actis:2007}. The latter method allows also to evaluate the contribution of
the light quarks vacuum polarization to the Bhabha scattering cross section; to
this purpose one has to convolute the kernel functions with the data
concerning the  cross section of the process $e^+ e^- \to$ hadrons.

\section{CONCLUSIONS \label{Conclusions}}

The numerical impact of the photonic and electron loop QED corrections to the
Bhabha scattering cross section at flavor factories was carefully examined in
\cite{BABAYAGA-1,CCCphipsi08}.
The authors were able to show that the event generator
BABAYAGA, based of the matching of exact NLO corrections with the Parton Shower
algorithm, has a theoretical accuracy of the order of $0.1 \%$.
A similar analysis of the heavy flavor NNLO corrections is not yet
available. However, it is possible to evaluate numerically
the NNLO heavy
flavor corrections to the Bhabha scattering cross section.
In Table~{\ref{tab1}} we show
the results of such an evaluation for  $\sqrt{s} = 1$ GeV and
for a scattering angle $\theta$ in the range $50^\circ < \theta < 130^\circ$
(see \cite{hfbhaJHEP} for details).
\begin{table*}
\begin{center}
{\rotatebox{90}{\makebox(0,0){\strut{} $\sqrt{s}=1$~GeV}}}
\hspace*{3mm}
\begin{tabular}{ c c c c c c c c }
\hline
$\theta$ & $phot$ ($10^{-4}$) & $e$ ($10^{-4}$) & $\mu$ ($10^{-4}$) & $c$ ($10^{-4}$)  &
$\tau$ ($10^{-4}$) & $b$ ($10^{-4}$) \cr
\hline
$50^{\circ}$ & 36.688225 & 17.341004 & 1.7972877 & 0.0622677 & 0.0264013 & 0.0010328 \cr
$70^{\circ}$ & 41.240039 & 19.438718 & 2.6504950 & 0.1086126 & 0.0465329 & 0.0018907 \cr
$90^{\circ}$ & 45.780639 & 21.463240 & 3.4581845 & 0.1321857 & 0.0576348 & 0.0024428 \cr
$110^{\circ}$ & 49.366078 & 23.099679 & 4.0922189 & 0.1098317 & 0.0495028 & 0.0022024 \cr
$130^{\circ}$ & 50.349342 & 23.847394 & 4.4392717 & 0.0549436 & 0.0273145 & 0.0013297 \cr
\hline
\end{tabular}
\caption{The second-order electron, muon,
$c$-quark, $\tau$-lepton, and $b$-quark QED contributions to the
differential cross section of Bhabha scattering at $\sqrt{s}=1$~GeV
in units of $10^{-4}$ of the Born cross section. \label{tab1}}
\end{center}
\end{table*}
It is possible to observe that the muon corrections are an
order of magnitude larger of the corrections involving heavier fermions while they  are
one order of magnitude smaller than the electron loop corrections and they reach 1/2
permille of the Born cross section at large scattering angles.

In conclusion, the calculation of the two-loop corrections to Bhabha scattering
in QED is now complete. The calculation of the heavy fermion NNLO corrections
allowed to remove the last piece of pure
theoretical uncertainty in the luminosity determination at low-energy
accelerators.

\end{document}